\documentclass[nofootinbib,showpacs,preprintnumbers,amsmath,amssymb,floatfix]
{revtex4}
\usepackage{epsf}

\begin{document}


\title{Infrared Evolution Equations: Method and Applications }

\vspace*{0.3 cm}

\author{B.I.~Ermolaev}
\affiliation{Ioffe Physico-Technical Institute, 194021
 St.Petersburg, Russia}
\author{M.~Greco}
\affiliation{Department of Physics and INFN, University Rome III,
Rome, Italy}
\author{S.I.~Troyan}
\affiliation{St.Petersburg Institute of Nuclear Physics, 188300
Gatchina, Russia}

\begin{abstract}
It is a brief review on composing and solving Infrared Evolution
Equations. They can be used in order to calculate amplitudes of
high-energy reactions in different kinematic regions in the
double-logarithmic approximation.
\end{abstract}

\pacs{12.38.Cy}

\maketitle

\section{Introduction}

Double-logarithmic (DL) contributions  are of a special interest
among radiative corrections. They are interesting in two aspects:
first, in every fixed order of the perturbation theories they are
the largest terms among the radiative corrections depending on the
total energy and second, they are easiest kind of the corrections
to sum up. DL corrections were discovered by V.V.~Sudakov in
Ref.~\cite{sud} in the QED context. He showed that DL terms appear
from integrations over soft, infrared (IR) -divergent momenta of
virtual photons. All-order resummation of such contributions led
to their exponentiations.

Next important step was done in Refs.~\cite{ggfl} where
calculation and summation of DL contributions was considered in a
systematic way. They found a complementary source of DL terms:
soft virtual fermions. This situation appears in the Regge
kinematics. The all-order resummations of DL contributions in the
Regge kinematic are quite involved and yield more complicated
expressions than the Sudakov exponentials. Nonetheless important
was the proof of the factorization of bremsstrahlung photons with
small $k_{\perp}$ in the high-energy hadronic reactions found in
Ref.~\cite{g} and often addressed as the Gribov's bremsstrahlung
theorem. This statement, suggested originally in the framework of
the phenomenological QED of hadrons was extended to QCD in
Refs.~\cite{efl}.

Calculation in the double-logarithmic approximation (DLA)
amplitudes of the fermion-antifermion annihilation in the Regge
forward and backward kinematics involves accounting for DL
contributions from soft quarks and soft gluons. These reactions in
QED and QCD have many common features. The $e^+e^-$ -annihilation
was studied in Refs.~\cite{ggfl}. The quark-aniquark annihilation
DLA was investigated in Ref.~\cite{kl}. The method of calculation
here was based on factorization of virtual quarks and gluons with
minimal $k_{\perp}$. Generally speaking, the results obtained in
Ref.~\cite{kl} could be obtained with the method of
Ref.~\cite{ggfl}, however the technique of calculations suggested
in Ref.~\cite{kl} was much more elegant and efficient. Although
Ref.~\cite{kl} is about quark scattering only, it contains almost
all technical ingredients necessary to compose Infrared Evolution
Equations for any of elastic scattering amplitudes.  Nevertheless
it could not directly be applied to inelastic processes involving
emission of soft particles. Such a generalization was obtained in
Refs.~\cite{efl,el}. The basic idea of the above-mentioned method
was suggested by  L.N.~Lipatov: to investigate evolution with
respect to the  infrared cut-off. The present, sounding naturally
term "Infrared Evolution Equations" (IREE) for this method was
suggested by M.~Krawczyk in Ref.~\cite{ek} where amplitudes for
the backward Compton scattering were calculated in DLA.

The aim of the present brief review is to show how to compose and
solve IREE for scattering amplitudes in different field theories
and kinematic regions. The paper is organized as follows: in
Sect.~II we consider composing IREE in the technically simplest
hard kinematics. In Sect.~III we consider composing IREE in the
forward kinematics and apply it to studying the structure function
$g_1$ of the polarized Deep-Inelastic scattering (DIS) at small
$x$. The point is that the commonly used theoretical instrument to
study $g_1$ is DGLAP \cite{dglap}. It collects logarithms of $Q^2$
to all orders in $\alpha_s$ but does not include the total
resummation of logarithms of $1/x$, though it is important at
small  $x$. Accounting for such a resummaton leads to the steep
rise of $g_1$ at the small-$x$ region. As is shown in Sect.~IV,
DGLAP lacks the resummaion but mimics it inexplicitly, through the
special choice of fits for the initial parton densities. Invoking
such peculiar fits together with DGLAP to describe $g_1$ at $x \ll
1$ led to various misconceptions in the literature. They are
enlisted and corrected in Sect.~V. The total resummaion of the
leading logarithms is essential in the region of small $x$. In the
opposite region of large $x$, DGLAP is quite efficient. It is
attractive to combine the resummation with DGLAP. The manual for
doing it is given in Sect.~VI. Finally, Sect.~VII is for
concluding remarks.

\section{IREE for scattering amplitudes in the hard kinematics}
From the technical point of view, the hard kinematics, where all
invariants are of the same order, is the easiest for analysis. For
the simplest, $2 \to 2$ -processes, the hard kinematics means that
the Mandelstamm variables $s,t,u$ obey
\begin{equation}\label{hard}
s \sim -t \sim -u~.
\end{equation} In  other words, the cmf
scattering angles $\theta \sim 1$ in the hard kinematics. This
kinematics is the easiest because the ladder Feynman graphs do not
yield DL contributions here and usually the total resummation of
DL contributions leads to multiplying the Born amplitude by
exponentials decreasing with the total energy. Let us begin with
composing and solving an IREE for the well-known object:
electromagnetic vertex $\Gamma_{\mu}$ of an elementary fermion
(lepton or quark). As is known,
\begin{equation}\label{v}
\Gamma_{\mu}  = \bar{u}(p_2)\big[ \gamma_{\mu} f(q^2) -
\frac{\sigma_{\mu\nu}q_{\nu}}{2m} g(q^2)\big]u(p_1)
\end{equation}
where $p_{1,2}$ are the initial and final momenta of the fermion,
$m$ stands for the fermion mass and the transfer momentum $q = p_2
- p_1$. Scalar functions $f$ and $g$ in Eq.~(\ref{v}) are called
form factors. Historically, DL contributions were  discovered by
V.~Sudakov when he studied the QED radiative corrections to the
form factor $f$ at $|q^2| \gg |p^2_{1,2}|$. Following him, let us
consider vertex $V_{\mu}$ at
\begin{equation}\label{sudkin}
|q^2| \gg p^2_1 = p^2_2 = m^2~
\end{equation}
i.e. we assume the fermion to be on--shell and account for DL
electromagnetic contributions. We will drop $m$ for the sake of
simplicity.
\subsection{IREE for the form factor $f(q^2)$ in QED}
\textbf{Step 1} is to introduce the infrared cut-off $\mu$ in the
transverse (with respect to the plane formed by momenta $p_{1,2}$)
momentum space for all virtual momenta $k_i$:
\begin{equation}\label{mu}
k_{i~\perp}  > \mu
\end{equation}
where $i  = 1,2,...$

\textbf{Step 2} is to look for the softest virtual particle among
soft external and virtual particles. The only option we have is
the softest virtual photon. Let denote its transverse momenta
$\equiv k_{\perp}$. By definition,
\begin{equation}\label{softest}
k_{\perp} = \min{k_{i~\perp}}~.
\end{equation}
\textbf{Step 3:} According to the Gribov  theorem, the propagator
of the softest photon can be factorized (i.e. it is attached to
the external lines in all possible ways) whereas $k_{\perp}$ acts
as a new cut-off for other integrations. Adding the Born
contribution $f^{Born}= 1$ we arrive at the IREE for $f$ in the
diagrammatic form. It is depicted in Fig.~1. IREE in the analytic
form are written in the gauge-invariant way, but their
diagrammatical writing depends on the gauge. In the present paper
we use the Feynman gauge.
\begin{figure}
\begin{center}
\begin{picture}(300,180)
\put(0,0){
\epsfbox{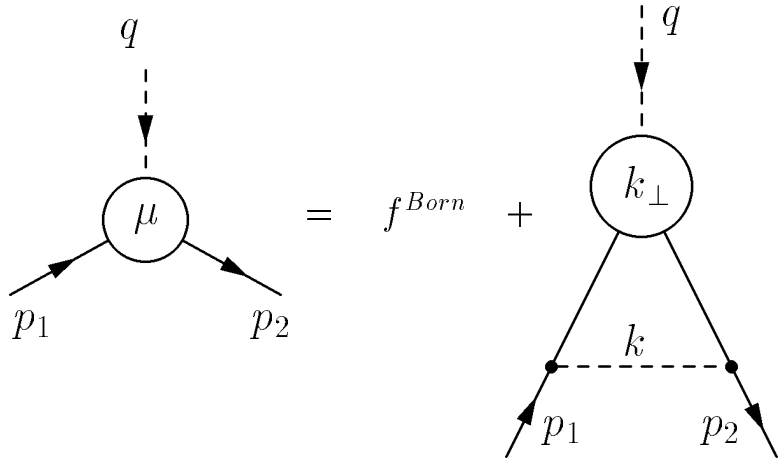} }
\end{picture}
\end{center}
\caption{\label{fig1}The IREE for the Sudakov form factor. The
letters in the blobs stand for IR cut-off.}
\end{figure}

Applying  to it the standard Feynman rules, we write it in the
analytic form:
\begin{equation}\label{eqf}
f(q^2,\mu^2) = f^{Born} -  \frac{e^2}{8\pi^2} \int \frac{d\alpha d
\beta d k_{\perp}^2~~~\Theta (k_{\perp}^2 - \mu^2)~~f(q^2,
k_{\perp}^2)}{(s\alpha\beta - k_{\perp}^2 +\imath
\epsilon)(-s\alpha + s\alpha\beta- k_{\perp}^2 + \imath
\epsilon)(s\beta + s\alpha\beta - k_{\perp}^2 + \imath \epsilon)}
\end{equation}
where we have used the Sudakov parametrization $k = \alpha p_2 +
\beta p_1 + k_{\perp}$ and denoted $s = -q^2\approx 2p_1p_2$. As
$f(q^2, k_{\perp}^2)$ does not depend on $\alpha$ and $\beta$, the
DL integration over them can be done with the standard way, so we
are left with a simple integral equation to solve:
\begin{equation}\label{eqfk}
f(q^2,\mu^2) = f^{Born} -  \frac{e^2}{8\pi^2} \int_{\mu^2}^{s}
\frac{d k_{\perp}^2}{k_{\perp}^2} \ln(s/k_{\perp}^2) f(q^2,
k_{\perp}^2)~.
\end{equation}
Differentiation of Eq.~(\ref{eqfk}) over $\mu^2$ (more exactly,
applying $-\mu^2
\partial/\partial \mu^2$) reduces it to a differential equation
\begin{equation}\label{difeqf}
\partial f/\partial (\ln(s/\mu^2)) = - (e^2/8\pi^2) \ln(s/\mu^2)  f
\end{equation}
with the obvious solution
\begin{equation}\label{solf}
f = f^{Born} \exp [-(\alpha/4\pi)\ln^2(q^2/m^2)]
\end{equation}
where we have replaced $\mu$ by $m$ and used $\alpha = e^2/4\pi$.
Eq.~(\ref{solf}) is the famous Sudakov exponential obtained in
Ref.~\cite{sud}.

\subsection{IREE for the form factor $g(q^2)$ in QED}

Repeating the same steps (see Ref.~\cite{et} for detail) leads to
a similar IREE for the form factor $g$:
\begin{equation}\label{eqgk}
g(q^2,m^2,\mu^2) = g^{Born}(s,m^2) -  \frac{e^2}{8\pi^2}
\int_{\mu^2}^{s} \frac{d k_{\perp}^2}{k_{\perp}^2}
\ln(s/k_{\perp}^2) g(q^2,m^2, k_{\perp}^2)~
\end{equation}
where $g^{Born}(s,m^2)  = -(m^2/s)(\alpha/\pi)\ln(s/m^2)$. Solving
this equation and putting $\mu = m$ in the answer leads to the
following relation between form factors $f$ and $g$:
\begin{equation}\label{fg}
g(s) = -2 \frac{\partial f}{\partial \rho}~,
\end{equation}
with $\rho = s/m^2$. Combining Eqs.~(\ref{solf},\ref{fg}) allows
to write a simple expression for the DL asymptotics of the vertex
$\Gamma_{\mu}$:
\begin{equation}\label{solv}
\Gamma_{\mu} =\bar{u}(p_2)\big[ \gamma_{\mu} +
\frac{\sigma_{\mu\nu}q_{\nu}}{m} \frac{\partial}{\partial
\rho}\big]u(p_1)\exp[-(\alpha/4\pi) \ln^2\rho]~.
\end{equation}

\subsection{$e^+e^-$ -annihilation into a
quark-antiquark pair}

Let us consider the $e^+e^-$ -annihilation into a quark $q(p_1)$
and $\bar{q}(p_2)$ at high energy when $2p_1p_2 \gg p^2_{1,2}$. We
consider the channel where the $e^+e^-$ -pair annihilates  into
one heavy photon which decays into the $q(p_1)~\bar{q}(p_2)$
-pair:
\begin{equation}\label{el}
e^+e^- \to \gamma^* \to  q(p_1)~\bar{q}(p_2)~.
\end{equation}
 We call this process elastic. In this case the most sizable
radiative corrections arise from the graphs where the quark and
antiquark exchange with gluons and these graphs look absolutely
similar to the graphs for the electromagnetic vertex
$\Gamma_{\mu}$ considered in the previous subsection. As a result,
accounting for the QCD radiative corrections in DLA to the elastic
form factors $f_q,~g_q$ of quarks can be obtained directly from
Eqs.~(\ref{solf},\ref{fg}) by replacement
\begin{equation}\label{gluons}
\alpha \to \alpha_s C_F,
\end{equation}
with $C_F = (N^2-1)/2N = 4/3$.

\subsection{$e^+e^-$ -annihilation into a
quark-antiquark pair and gluons}

In addition to the elastic annihilation (\ref{el}), the final
state can include gluons:
\begin{equation}\label{inel}
e^+e^- \to \gamma^* \to  q(p_1)~\bar{q}(p_2)  + g(k_1),..g(k_n)~.
\end{equation}
We call this process the inelastic annihilation. The QED radiative
corrections to the inelastic annihilation (\ref{inel}) in DLA are
absolutely the same as the corrections to the elastic
annihilation. On the contrary, the QCD corrections account for
gluon exchanges between all final particles. This makes composing
the IREE for the inelastic annihilation be more involved (see
Ref.~\cite{efl}). The difference to the considered elastic case
appears at \textbf{Step 2}: look for the softest virtual particle
among soft external and virtual particles. Indeed, now the softest
particle can be both a virtual gluon and an emitted  gluon. For
the sake of simplicity let us discuss the 3-particle final state,
i.e. the process
\begin{equation}\label{inel1}
e^+e^- \to \gamma^* \to  q(p_1)~\bar{q}(p_2) + g(k_1)~.
\end{equation}
The main ingredient of the scattering amplitude of this process is
the new electromagnetic vertex $\Gamma^{(1)}_{\mu}$ of the quark.
In DLA, it is parameterized  by new form factors $F^{(1)}$ and
$G^{(1)}$
\begin{equation}\label{vinel}
\Gamma_{\mu}  = B_1(k_1) \bar{u}(p_2)\big[ \gamma_{\mu}
F^{(1)}(q,k_1) - \frac{\sigma_{\mu\nu}q_{\nu}}{2m}
G^{(1)}(q,k_1)\big]u(p_1)
\end{equation}
where (1) corresponds to the number of emitted gluons, $q = p_1 +
p_2$ and $l$ is the polarization vector of the emitted gluon. The
bremsstrahlung factor $B_1$ in Eq.~(\ref{vinel}) at high energies
is expressed through  $k_{1~\perp}$:
\begin{equation}\label{b1}
B_1 = \Big(\frac{p_2l}{p_2k_1} - \frac{p_1l}{p_1k_1}\Big) \approx
\frac{2}{k_{1\perp}}~.
\end{equation}

We call $F^{(n)}, G^{(n)}$ inelastic form factors. Let us start
composing the IREE for $F^{(1)}$. \textbf{Step 1} is the same like
in the previous case. \textbf{Step 2} opens more options. Let us
first choose the softest gluon among virtual gluons and denote its
transverse momentum  $k_{\perp}$  The integration over $k_{\perp}$
runs from $\mu$ to $s$. As $\mu < k_{1~\perp} < s$, we have two
regions to consider: Region $\emph{D}_1$ were
\begin{equation}\label{d1}
 ~~~~\mu < k_{1\perp} < k_{\perp}  < \sqrt{s}
\end{equation}
and Region $\emph{D}_2$ were
\begin{equation}\label{d2}
 ~~~~\mu < k_{\perp} < k_{1\perp}  < \sqrt{s}
\end{equation}
Obviously, the softest particle in  Region $\emph{D}_1$ is the
emitted gluon, so it can be factorized as depicted in graphs
(b,b') of Fig.~2.
\begin{figure}
\begin{center}
\begin{picture}(400,410)
\put(0,0){
\epsfbox{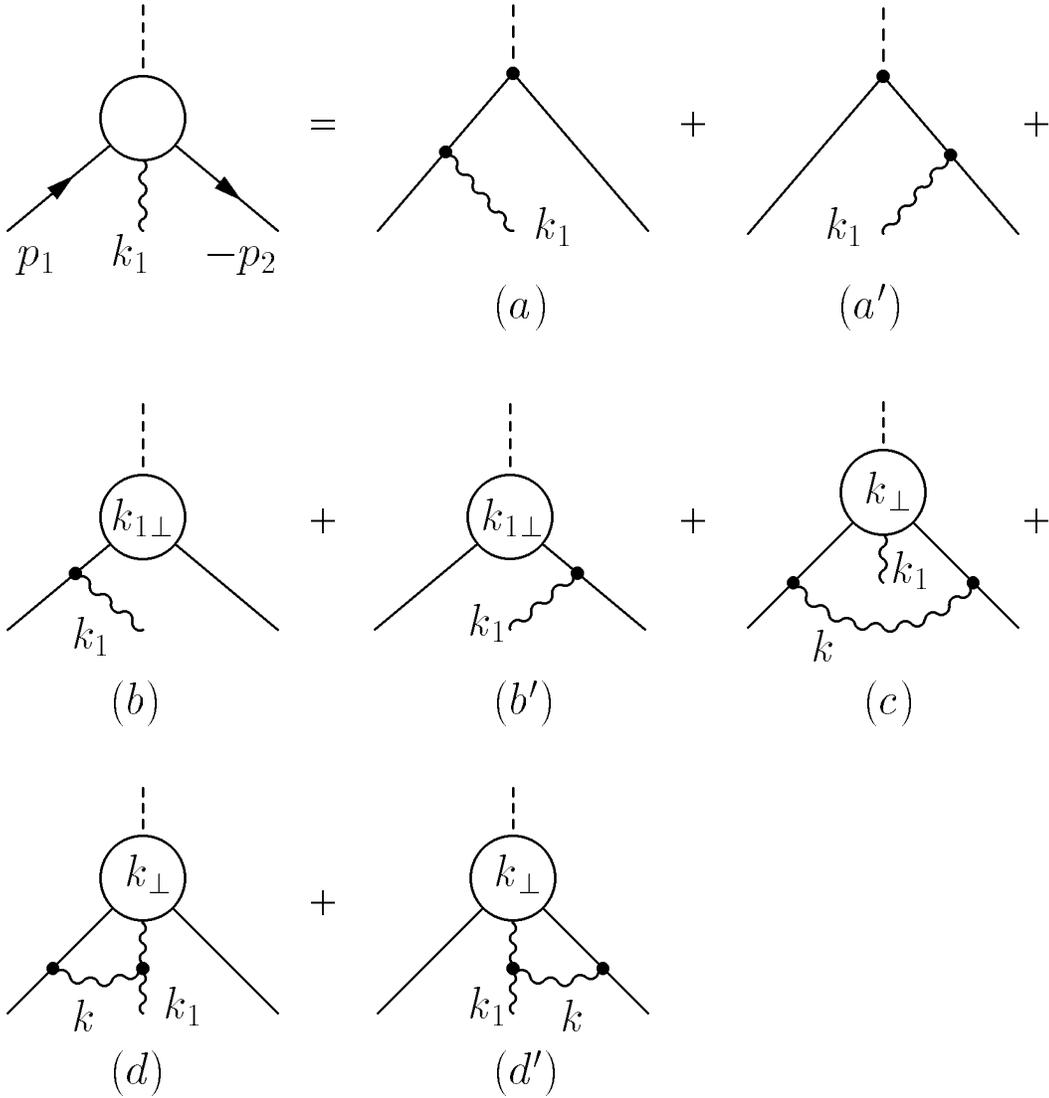} }
\end{picture}
\end{center}
\caption{\label{fig2}The IREE for the inelastic quark form
factor.}
\end{figure}

On the contrary, the virtual gluon is the softest  in Region
$\emph{D}_2$ were its propagator is factorized as shown in graphs
(c,d,d') of Fig.~2. Adding the Born contribution (graphs (a,a') in
Fig.~2) completes the IREE for $F^{(1)}$ depicted in Fig.~2.
Graphs (a-b') do not depend on $\mu$ and vanish when
differentiated with respect to $\mu$. Blobs in graphs  (c-d') do
not depend on the longitudinal Sudakov variables, so integrations
over $\alpha,\beta$ can be done like in the first loop. After that
the differential IREE for $F^{(1)}$ is
\begin{equation}
\label{eqf1} -\mu^2 \frac{\partial F^{(1)}}{\partial \mu^2} =
-\frac{\alpha_s}{2\pi} \Big[ C_F \ln \Big(\frac{s}{\mu^2} \Big) +
\frac{N}{2}\ln \Big(\frac{2p_2 k_1}{\mu^2} \Big) + \frac{N}{2}\ln
\Big(\frac{2p_1k_1}{\mu^2} \Big) \Big]F^{(1)}~.
\end{equation}
Solving Eq.~(\ref{eqf1}) and using that $(2p_1k_1)(2p_2k_1) = s
k^2_{1 \perp}$ leads to the expression
\begin{equation}\label{f1}
F^{(1)} = \exp \Big(-\frac{\alpha_s}{4\pi}\Big[C_F \ln^2
\Big(\frac{s}{\mu^2}\Big) + \frac{N}{2}\ln^2
\Big(\frac{k^2_{1\perp}}{\mu^2} \Big) \Big]\Big)
\end{equation}
suggested in Ref.~\cite{ef} and proved in Ref.~\cite{efl} for any
$n$. The IREE for the form factor $G^{(n)}$ was obtained and
solved in Ref.~\cite{et}.  It was shown that
\begin{equation}\label{f1g1}
G^{(n)} =  -2 \partial F^{(n)}/\partial \rho~.
\end{equation}

\subsection{Exponentiation of Sudakov electroweak double-logarithmic contributions}

The IREE -method was applied in Ref.~\cite{flmm} to prove
exponentiation of DL correction to the electroweak (EW) reactions
in the hard kinematics. There is an essential technical difference
between the theories  with the exact gauge symmetry (QED and QCD)
and the EW interactions theory with the broken $SU(2)\otimes U(1)$
gauge symmetry: only DL contributions from virtual photons yield
IR singularities needed to be regulated with the cut-off $\mu$
whereas DL contributions involving $W$ and $Z$ -bosons are IR
stable because the boson masses $M_W$ and $M_Z$ act as IR
regulators. In Ref.~\cite{flmm} the difference between $M_W$ and
$M_Z$ was neglected and the parameter
\begin{equation}\label{m}
M \gtrsim M_W \approx M_Z
\end{equation}
was introduced,  in addition to $\mu$, as the second IR cut-off.
It allowed to drop masses $M_{W,Z}$. The IREE with two IR cut-offs
was composed quite similarly to Eq.~(\ref{eqf}), with factorizing
one by one the softest virtual photon, $Z$-boson and $W$-boson. As
a result the EW Sudakov form factor $F_{EW}$ is
\begin{equation}\label{few}
F_{EW} = \exp\Big( -\frac{\alpha
(Q^2_1+Q^2_2)}{8\pi}\ln^2(s/\mu^2) -\Big[\frac{g^2
C_F^{SU(2)}}{16\pi^2} + \frac{g'^2}{16\pi^2}\frac{(Y^2_1 +
Y^2_2)}{4} - \frac{\alpha (Q^2_1+Q^2_2)}{8\pi}\Big]\ln^2(s/M^2)
\Big)
\end{equation}
where $Q_{1,2}$ are the electric charges of the initial and final
fermion (with $W$ -exchanges accounted, they may be different),
$Y_{1,2}$ are their hyper-charges and $C_F^{SU(2)} = (N^2-1)/2N$,
with $N=2$. We have used in Eq.~(\ref{few}) the standard notations
$g$ and $g'$ for the $SU(2)$ and $U(1)$ -EW couplings. The
structure of the exponent in Eq.~(\ref{few}) is quite clear: the
first, $\mu$ -dependent term comes from the factorization of soft
photons like the exponent in Eq.~(\ref{solf}) while other terms
correspond to the $W$ and $Z$ -factorization; the factor in the
squared brackets is the sum of the $SU(2)$ and U(1) Casimirs, with
the photon Casimir being subtracted to avoid the double counting.
In the limit $\mu = M$ the group factor in the exponent is just
the Casimir of $SU(2)\otimes U(1)$.

\section{Application of IREE to the polarized Deep-Inelastic Scattering}

Cross-sections of the polarized DIS are described by the structure
functions $g_{1,2}$. They appear from the standard parametrization
of the spin-dependent part $W_{\mu\nu}$ of the hadronic tensor:
\begin{equation}\label{w}
W_{\mu\nu} = \imath
\epsilon_{\mu\nu\lambda\rho}q_{\lambda}\frac{m}{pq} \Big[
S_{\rho}g_1(x,Q^2) + \Big(S_{\rho} - p_{\rho} \frac{Sq}{pq}
\Big)g_2(x,Q^2) \Big]
\end{equation}
where  $p$, $m$  and $S$ are the momentum, mass and spin of the
incoming hadron;  $q$ is the virtual photon momentum; $Q^2 =
-q^2$; $x = Q^2/2pq$. Obviously, $Q^2 \geqslant 0$ and $0
\leqslant x \leqslant 1$.

Unfortunately, $g_{1,2}$ cannot be calculated in a straightforward
model-independent way because it would involve QCD at long
distances. To avoid this problem, $W_{\mu\nu}$ is regarded as a
convolution of $\Phi_{q,g}$ - probabilities to find a polarized
quark or gluon and the partonic tensors
$\tilde{W}^{(q,g)}_{\mu\nu}$ parameterized identically to
Eq.~(\ref{w}). In this approach $\tilde{W}^{(q,g)}_{\mu\nu}$
involve only QCD at short distances, i.e. the Perturbative QCD
while long-distance effects are accumulated in $\Phi_{q,g}$. As
$\Phi_{q,g}$ are unknown, they are mimicked by the initial quark
and gluon densities $\delta q,~\delta g$. They are fixed
aposteriori from phenomenological considerations. So, the standard
description of DIS is:
\begin{equation}\label{wc}
W_{\mu\nu} \approx W_{\mu\nu}^{(q)}\otimes \delta q +
W_{\mu\nu}^{(g)}\otimes \delta g~.
\end{equation}

The standard theoretical instrument to calculate $g_1$ is
DGLAP\cite{dglap} complemented with standard fits\cite{fits} for
$\delta q,~\delta g$. We call it \textbf{Standard Approach} (SA).
In this approach

\begin{equation}\label{g1sa}
g_1(x,Q^2) = C_q (x/z)\otimes \Delta  q(z, Q^2)+C_g (x/z)\otimes
\Delta  g(z, Q^2)
\end{equation}
where $C{q,g}$ are coefficient functions and $\Delta  q(z,
Q^2),~\Delta  g(z, Q^2)$ are called the evolved (with respect to
$Q^2$)quark and gluon distributions. They are found as solutions
to DGLAP evolution equations
\begin{equation}\label{dglap}
\frac{d \Delta q}{d \ln Q^2} = \frac{\alpha_s(Q^2)}{2\pi}
\big[P_{qq}\Delta q + P_{qg}\Delta g \big],~~\frac{d \Delta g}{d
\ln Q^2} = \frac{\alpha_s(Q^2)}{2\pi} \big[P_{gq}\Delta q +
P_{gg}\Delta g \big]
\end{equation}
where $P_{ab}$ are the splitting functions. The Mellin transforms
$\gamma_{ab}$ of $P_{ab}$ are called the DGLAP anomalous
dimensions. They are known in the leading order (LO) where they
are $\sim \alpha_s$ and in the next-to-leading order (NLO), i.e.
$\sim \alpha_s^2$. Similarly, $C_{q,g}$ are known in LO and NLO.
Details on this topic can be found in the literature (e.g. see a
review \cite{vn}). Structure function $g_1$ has the flavor singlet
and non-singlet components, $g_1^S$ and $g_1^{NS}$. Expressions
for $g_1^{NS}$ are simpler, so we will use mostly them in the
present paper when possible. It is convenient to write $g_1$ in
the form of the Mellin integral. In particular,
\begin{equation}
\label{nsdglap} g_1^{NS~DGLAP}(x, Q^2) = (e^2_q/2) \int_{-\imath
\infty}^{\imath \infty} \frac{d \omega}{2\pi\imath }\Big(
\frac{1}{x} \Big)^{\omega} C_{NS}(\omega) \delta q(\omega) \exp
\Big[\int_{\mu^2}^{Q^2}\frac{d
k^2_{\perp}}{k^2_{\perp}}\gamma_{NS}(\omega,
\alpha_s(k^2_{\perp}))\Big]
\end{equation}
where $\mu^2$ is the starting point of the $Q^2$ -evolution;
$C_{NS}$ and $\gamma_{NS}$ are the non-singlet coefficient
function and anomalous dimension. In LO
\begin{eqnarray}\label{lo}
\gamma_{NS}(\omega, Q^2) =
\frac{\alpha_s(Q^2)C_F}{2\pi}\Big[\frac{1}{\omega(1+\omega)}+\frac{3}{2}
+ S_2(\omega)\Big], \\
\nonumber~~C_{NS}^{LO}(\omega) = 1
+\frac{\alpha_s(Q^2)C_F}{2\pi}\Big[ \frac{1}{\omega^2} +
 \frac{1}{2\omega}+\frac{1}{2\omega +1} -\frac{9}{2}
+\Big(\frac{3}{2}-\frac{1}{\omega(1+\omega)}\Big)\Big(S_1(\omega)+S^2_1(\omega)-S_2(\omega)
\Big)\Big]
\end{eqnarray}
with $S_r(\omega)=\sum_{j=1}^{\omega} 1/j^r$~. The initial quark
and gluon densities in Eq.~(\ref{nsdglap}) are defined through
fitting experimental data. For example, the fit for $\delta q$
taken from the  first paper in Ref.~\cite{fits} is
\begin{equation}
\label{fita} \delta q(x) = N x^{- \alpha} \Big[(1 -x)^{\beta}(1 +
\gamma x^{\delta})\Big],
\end{equation}
with $N$ being the normalization, $\alpha = 0.576$, $\beta =
2.67$, $\gamma = 34.36$ and $\delta = 0.75$.

DGLAP equations were suggested for describing DIS in the region
\begin{equation}\label{dglapreg}
x\lesssim 1,~~~~~~~~~Q^2 \gg \mu^2
\end{equation}
($\mu$ stands for a mass scale, $\mu \gg \Lambda_{QCD}$)  and
there is absolutely no theoretical grounds to apply them in the
small-$x$ region, however being complemented with the standard
fits they are commonly used at small $x$. It is known that SA
provide a good agreement with available experimental data but the
price is invoking a good deal of phenomenological parameters. The
point is that DGLAP, summing up leading $\ln^k Q^2$ to all orders
in $\alpha_s$, cannot do the same with leading $\ln^k(1/x)$. The
later is not important in the region (\ref{dglapreg}) where
$\ln^k(1/x) \ll 1$ but becomes a serious drawback of the method at
small $x$. The total resummation of DL contributions to $g_1$ in
 the region
\begin{equation}\label{smx}
x\ll 1,~~~~~~~~~Q^2 \gg \mu^2
\end{equation}
was done in Refs.~\cite{ber}. The weakest point in those papers
was keeping $\alpha_s$ as a parameter, i.e. fixed at an unknown
scale. Accounting for the most important part of
single-logarithmic contributions, including the running coupling
effects were done in Refs.~\cite{egt}. In these papers $\mu^2$ was
treated as the starting point of the $Q^2$ -evolution and as the
IR cut-off at the same time. The structure function $g_1$ was
calculated with composing and solving IREE in the following way.

It is convenient to compose IREE not for $g_1$ but for forward
(with $|t| \lesssim \mu^2$) Compton amplitude $M$ related to $g_1$
as follows:
\begin{equation}\label{mg}
g_1 = \frac{1}{\pi}  \Im M~.
\end{equation}
It is also convenient to use for amplitude $M$ the asymptotic form
of the Sommerfeld-Watson transform:
\begin{equation}\label{mellin}
M = \int_{-\imath \infty}^{\imath \infty}\frac{d \omega}{2\pi
\imath}\Big(\frac{s}{\mu^2}\Big)^\omega \xi^{(-)}(\omega)F(\omega,
Q^2/\mu^2)
\end{equation}
where $\xi^{(-)}(\omega) = [e^{-\imath \pi\omega}  -1]/2  \approx
-\imath \pi\omega/2 $ is the signature factor. The transform of
Eq.~(\ref{mellin}) and is often addressed as the Mellin transform
but one should remember that it coincides with the Mellin
transform only partly.  IREE for Mellin amplitudes $F(\omega,
Q^2)$ look quite simple.

 For example, the IREE for
the non-singlet Mellin amplitude $F^{NS}$ related to $g_1^{NS}$ by
Eqs.~(\ref{mg},\ref{mellin}) is depicted in Fig.~3.
\begin{figure}
\begin{center}
\begin{picture}(380,160)
\put(0,0){
\epsfbox{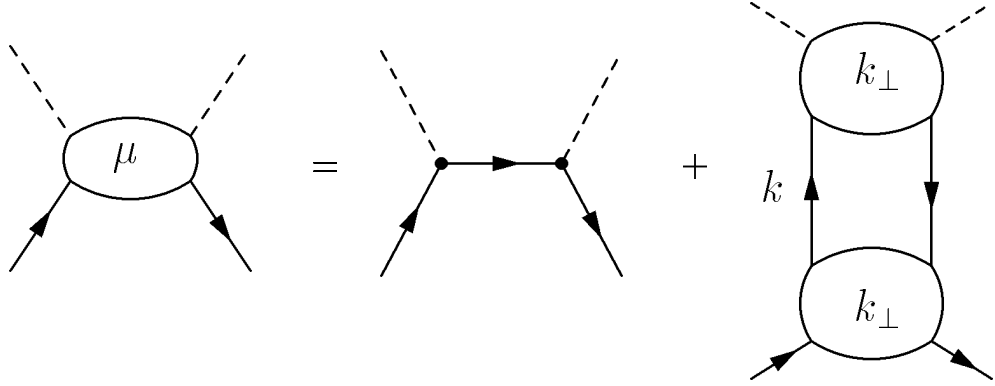} }
\end{picture}
\end{center}
\caption{\label{fig3}The IREE for the non-singlet component of the
spin structure function $g_1$.}
\end{figure}
In the Mellin space it takes the simple form:
\begin{equation}\label{eqfns}
[\omega + \partial/\partial y]F^{NS} = (1+\omega/2)H_{NS}F^{NS}~
\end{equation}
where $y = \ln(Q^2/\mu^2)$. Eq.~(\ref{eqfns})  involves a new
object (the lowest blob in the last term in Fig.~3): the
non-singlet anomalous dimension $H_{NS}$ accounting for the total
resummaton of leading logarithms of $1/x$. Like in DGLAP, the
anomalous dimension does not depend on $Q^2$ but, in contrast to
DGLAP, $H_{NS}$ can be found with the same method. The IREE for it
is algebraic:
\begin{equation}\label{eqhns}
\omega H_{NS}  = A(\omega)C_F/8\pi^2 + (1 + \omega/2)H_{NS}^2 +
D(\omega)/8\pi^2~.
\end{equation}
The system of Eqs.~(\ref{eqfns},\ref{eqhns}) can be easily solved
but before doing it let us comment on them. The left-hand sides of
Eqs.~(\ref{eqfns},\ref{eqhns}) are obtained with applying the
operator $-\mu^2
\partial/\partial \mu^2$  to Eq.~(\ref{mellin}). The Born contribution in Fig.~3
does not depend on $\mu$ and therefore vanishes. The last term in
Fig.~3 (the rhs of Eq.~(\ref{eqfns})) is the result of a new, $t$
-channel factorization which does not exist in the hard kinematics
defined in Eq.~(\ref{hard}). In order to compose the IREE for the
Compton amplitude $M$, in accordance with the prescription in the
previous section we should first introduce the cut-off $\mu$. Then
\textbf{Step 2} is to tag the softest particles. In the case under
discussion we do not have soft external particles. Had the softest
particle been a gluon, it could be factorized in the same way like
in Sect.~II. However, the only option now is to  attach the
softest propagator to the external quark lines and get $\ln
(t/\mu^2) = 0$ from integration over $\beta$ (cf
Eq.~(\ref{eqfk})). So, the softest gluon does not yield DL
contributions. The other option is to find a softest quark. The
softest $t$ -channel quark pair factorizes amplitude $M$ into two
amplitudes (the last term in Fig.~3) and yield DL contributions.
 The IREE for $H_{NS}$ is different:

(\textbf{i}) $H_{NS}$ does not depend on $Q^2$, so there is not a
derivative in the lhs of  Eq.~(\ref{eqfns}).

(\textbf{ii}) The Born term depends on $\mu$ and contributes to
the IREE (term $A$ in Eq.~(\ref{eqfns}))).

(\textbf{iii}) As all external particles now are quarks, the
softest virtual particle can be both a quark and gluon. The case
when it is the $t$ -channel quark pair, corresponds to the
quadratic term in the rhs of Eq.~(\ref{eqfns}). The case of the
softest gluon yields the term $D$, with

\begin{equation}
\label{d} D(\omega) = \frac{2C_F}{b^2 N} \int_0^{\infty} d \eta
e^{-\omega \eta} \ln \big( \frac{\rho + \eta}{\eta}\big) \Big[
\frac{\rho + \eta}{(\rho + \eta)^2 + \pi^2} - \frac{1}{\eta}\Big]
\end{equation}
where $b = (33 - 2n_f)/12\pi$ and $\eta =
\ln(\mu^2/\Lambda^2_{QCD})$.

 The term $A$ in Eq.~(\ref{eqfns}) stands
instead of $\alpha_s$. The point is that the standard
parametrization  $\alpha_s = \alpha_s(Q^2)$ cannot be used at $x
\ll 1$ and should be changed (see Ref.~\cite{egta} for detail). It
leads to the replacement $\alpha_s$ by
\begin{equation}\label{a}
A(\omega) = \frac{1}{b} \Big[ \frac{\eta}{\eta^2 + \pi^2} -
\int_{0}^{\infty} \frac{d \rho e^{- \omega \rho}}{(\rho + \eta)^2
+ \pi^2} \Big]~.
\end{equation}

Having solved Eqs.~(\ref{eqfns},\ref{eqhns}), we arrive at the
following expression for $g_1^{NS}$ in the region (\ref{smx}):
\begin{equation}
\label{fnsint} g_1^{NS}(x, Q^2) = (e^2_q/2) \int_{-\imath
\infty}^{\imath \infty} \frac{d \omega}{2\pi\imath}(1/x)^{\omega}
C_{NS}(\omega) \delta q(\omega) \exp\big( H_{NS}(\omega) y\big)
\end{equation}
where the coefficient function $C_{NS}(\omega)$ is expressed
through $H_{NS}(\omega)$:
\begin{equation}
\label{cns} C_{NS}(\omega) =\frac{\omega}{\omega -
H_{NS}(\omega)}~
\end{equation}
and $H_{NS}(\omega)$ is the solution of algebraic equation
(\ref{hns}):

\begin{equation}
\label{hns} H_{NS} = (1/2) \Big[\omega - \sqrt{\omega^2 -
B(\omega)} \Big]
\end{equation}
where
\begin{equation}
\label{b} B(\omega) = (4\pi C_F (1 +  \omega/2) A(\omega) +
D(\omega))/ (2 \pi^2)~.
\end{equation}
 It is shown in Ref.~\cite{smq}
 that the expression for $g_1$
in the region

\begin{equation}\label{smxq}
x\ll 1,~~~~~~~~~Q^2 \lesssim \mu^2
\end{equation}
can be obtained from the expressions obtained in Refs.~\cite{egt}
for $g_1$ in region (\ref{smx})  by the shift
\begin{equation}\label{shift}
Q^2 \to Q^2 + \mu^2_0~
\end{equation}
where $\mu_0 = 1$~GeV for the non-singlet $g_1$ and $\mu_0 =
5.5$~GeV for the singlet.

\section{Comparison of expressions (\ref{nsdglap}) and (\ref{fnsint}) for $g_1^{NS}$}

Eqs.~(\ref{nsdglap}) and (\ref{fnsint}) read that the non-singlet
$g_1$ is obtained from $\delta q$ with evolving it with respect to
$x$ (using the coefficient function) and with respect to $Q^2$
(using the anomalous dimension).  Numerical comparison of
Eqs.~(\ref{nsdglap}) and (\ref{fnsint}) can be done when $\delta
q$ is specified.

\subsection{Comparison of small-$x$ asymptotics, neglecting the impact of $\delta q$ }

In the first place let us compare the small-$x$ asymptotics of for
$g_1^{NS~DGLAP}$ and $g_1^{NS}$, assuming that $\delta  q$ does
not affect them. In other words, we compare the differencee in the
$x$-evolution at $x \to 0$. Applying the saddle-point method to
Eqs.~(\ref{nsdglap}) and (\ref{fnsint}) leads to the following
expressions:

\begin{equation}\label{asdglap}
g_1^{NS~DGLAP} \sim \exp
\Big[\sqrt{\ln(1/x)\ln\ln(Q^2/\Lambda^2_{QCD})}\Big]
\end{equation}
and
\begin{equation}\label{as}
g_1^{NS} \sim (1/x)^{\Delta_{NS}} (Q^2/\mu^2)^{\Delta_{NS}/2}
\end{equation}
where ${\Delta_{NS}}=0.42$ is the non-singlet
intercept\footnote{The singlet intercept is much greater:
$\Delta_S = 0.86.$}. Expression (\ref{asdglap}) is the well-known
DGLAP asymptotics. Obviously, the asymptotics (\ref{as}) is much
steeper than the DGLAP asymptotics (\ref{nsdglap}).

\subsection{Numerical comparison between Eqs.~(\ref{nsdglap}) and (\ref{fnsint}),
neglecting the impact of $\delta q$}

A comparison between Eqs.~(\ref{nsdglap}) and (\ref{fnsint})
strongly depends on the choice of $\delta q$ but also depends on
the difference between the coefficient functions and anomalous
dimensions. To clarify the latter we choose the simplest form of
$\delta q$:
\begin{equation}\label{quark}
\delta q(\omega)  = N_q~.
\end{equation}
It corresponds to the evolution from the bare quark where $\delta
q(x) = N_q \delta(1-\mu^2/s)$. Numerical results for $R =
[g_1^{NS}-g_1^{NS~DGLAP}]/g_1^{NS~DGLAP}$ with $\delta q$ chosen
by Eq.~(\ref{quark}) manifest (see Ref.~\cite{egtfit} for detail)
that $R$ increases when $x$ is decreases. In particular, $R > 0.3$
at $x \lesssim 0.05$. It means that the total resummation of
leading $\ln^k(1/x)$ cannot be neglected at $x \lesssim 0.05$ and
DGLAP cannot be used beyond $x \approx 0.05$. On the other hand,
it is well--known that Standard Approach based on DGLAP works well
at $x \ll 0.05$. To solve this puzzle, we have to consider  the
standard fit for $\delta q$ in more detail.

\subsection{Analysis of the standard fits for $\delta q$}

There are known different fits for $\delta q$. We consider the fit
of  Eq.~(\ref{fita}). Obviously, in the $\omega$ -space
Eq.~(\ref{fita}) is a sum of pole contributions:
\begin{equation}
\label{fitaomega} \delta q(\omega) = N \eta \Big[ (\omega -
\alpha)^{-1} + \sum m_k (\omega + \lambda_k)^{-1}\Big],
\end{equation}
with $\lambda_{k} > 0$, so that the first term in
Eq.~(\ref{fitaomega}) corresponds to the singular term
$x^{-\alpha}$ of Eq.~(\ref{fita}) and therefore the small-$x$
asymptotics of $f_{DGLAP}$  is given by the leading singularity
$\omega = \alpha = 0.57$ of the integrand in Eq.~(\ref{fitaomega})
so that the asymptotics of $g_1^{NS~DGLAP}(x, Q^2)$ is not given
by the classic exponential of Eq.~(\ref{asdglap}) but actually is
the Regge-like:
\begin{equation}\label{assa}
g_1^{NS~DGLAP} \sim
C(\alpha)(1/x)^{\alpha}\Big(\ln(Q^2/\Lambda^2)/
\ln(\mu^2/\Lambda^2)\Big)^{\gamma(\alpha)/b},
\end{equation}
with $b = (33 - 2n_f)/12\pi$. Comparison of Eq.~(\ref{as}) and
Eq.~(\ref{assa}) demonstrates that both DGLAP and our approach
lead to the Regge behavior of $g_1$, though the DGLAP prediction
is more singular than ours. Then, they predict different $Q^2$
-behavior. However, it is important that our intercept
$\Delta_{NS}$ is obtained by the total resummation of the leading
logarithmic contributions and without assuming singular fits for
$\delta q$ whereas the SA intercept $\alpha$ in
Eq.~(\ref{asdglap}) is generated by the phenomenological factor
$x^{-0.57}$ of Eq.~(\ref{fita}) which makes the structure
functions grow when $x$ decreases and mimics in fact the total
resummation\footnote{We remind that our estimates for the
intercepts $\Delta_{NS}, \Delta_{S}$ were confirmed (see
Refs.~\cite{kat}) by analysis  of the experimental data}. In other
words, the role of the higher-loop radiative corrections on the
small-$x$ behavior of the non-singlets is, actually, incorporated
into SA phenomenologically, through the initial parton densities
fits. It means that the singular factors can be dropped from such
fits
 when the coefficient functions account for the total resummation
of the leading logarithms and therefore fits for $\delta q$ become
regular in $x$ in this case. They also can be simplified. Indeed,
if $x$ in the regular part $N \Big[(1 -x)^{\beta}(1 + \gamma
x^{\delta})\Big]$ of the fit (\ref{fita}) is not large, all $x$
-dependent terms can be neglected. So, instead of the rather
complicated expression of Eq.~(\ref{fita}), $\delta q$ can be
approximated by a constant or a linear form
\begin{equation}\label{fit}
\delta q(x) = N(1 + ax)~.
\end{equation}
with 2 phenomenological parameters instead of 5 in
Eq.~(\ref{fita}).

\section{Correcting misconceptions}

The total resummation of $\ln^k(1/x)$ allows to correct several
misconceptions popular in the literature. We list and correct them
below.\\

\textbf{Misconception 1}: Impact of non-leading perturbative and
non-perturbative contributions on the intercepts of $g_1$ is
large.

\textbf{Actually:} Confronting our results and the estimates of
the intercepts in Refs.~\cite{kat} obtained from fitting available
experimental data manifests that the total contribution of
non-leading perturbaive and non-perturbative contributions to the
intercepts is very small, so the main impact on the intercepts is
brought by the leading logarithms. \\

\textbf{Misconception 2}: Intercepts of $g_1$ should depend on
$Q^2$ through  the parametrization of the QCD coupling $\alpha_s =
\alpha_(Q^2)$.

\textbf{Actually:} This is groundless from the theoretical point
of view and appears only if the the parametrization of the QCD
coupling $\alpha_s = \alpha_(k_{\perp}^2)$ is kept in all ladder
rungs. It is shown in Ref.~\cite{egta} that this parametrization
cannot be used at small $x$ and should be replaced by the
parametrization of Eq.~(\ref{a}).\\

\textbf{Misconception 3}: Initial densities $\delta q(x)$ and
$\delta g(x)$ are singular but they are defined at $x$ not too
small. Later, being convoluted with the coefficient functions,
they become less singular.

\textbf{Actually:} It is absolutely wrong: Eq.~(\ref{fitaomega})
proves that the pole singularity $x^{-\alpha}$ in the fits does
not become weaker with the $x$-evolution.  \\

 \textbf{Misconception 4}: Fits for the
initial parton densities are complicated because they mimic
unknown non-perturbative contributions.

\textbf{Actually:} Our results demonstrate that the singular
factors in the fits mimic the total resummation of $\ln^k(1/x)$
and can be dropped when the resummation is accounted for. In the
regular part of the fits the $x$ -dependence is essential for
large $x$ only, so impact of non-perturbative contributions is
weak at the small-$x$ region.  \\

\textbf{Misconception 5}: Total resummations of $\ln^k(1/x)$ may
become of some importance at extremely small $x$ but not for $x$
available presently and in a forthcoming future.

\textbf{Actually:} The efficiency of SA in the available small-$x$
range is based on exploiting the singular factors in the standard
fits to mimic the resummations. So, the resummations have always
been used in SA at small $x$ in an inexplicit way, through the
fits, but without being aware of it.

\section{Combining the total resummation and  DGLAP}

The total resummaton of leading logarithms of $x$ considered in
Sect.~IV is essential at small-$x$.  When $x \sim 1$, all terms
$\sim \ln^k(1/x)$ in the coefficient functions and anomalous
dimensions cannot have a big impact compared to other terms. DGLAP
accounts for those terms. It makes DGLAP be more precise at large
$x$ than our approach. So, there appears an obvious appeal to
combine the DGLAP coefficient functions and anomalous dimensions
with our expressions in order to obtain an approach equally good
in the whole range of $x:~0<x<1$. The prescription for such
combining was suggested in Ref.~\cite{egtfit}. Let us, for the
sake of simplicity, consider here combining the total resummation
and LO DGLAP. The generalization to NLO DGLAP can be done quite
similarly. The prescription consists of the following points:

\textbf{Step A:} Take Eqs.~(\ref{lo}) and  replace $\alpha_s$ by
$A$ of Eq.~(\ref{a}), converting $\gamma_{NS}$ into
$\tilde{\gamma}_{NS}$ and $C_{NS}^{LO}$ into
$\tilde{C}_{NS}^{LO}$.

\textbf{Step B:} Sum  up the obtained expressions and
Eqs.~(\ref{cns},\ref{hns}):
\begin{equation}\label{sum}
\tilde{c}_{NS} = \tilde{C}_{NS}^{LO} + H_{S},~~~\tilde{h}_{NS} =
\tilde{\gamma}_{NS} + H_{NS}~.
\end{equation}
New expressions $\tilde{c}_{NS}, \tilde{h}_{NS}$ combine the total
resummation and DGLAP but they obviously contain the double
counting: some of the first--loop contributions are present both
in Eqs.~(\ref{lo}) and in Eqs.~(\ref{cns},\ref{hns}). To avoid the
double counting, let us expend Eqs.~(\ref{cns},\ref{hns}) into
series and retain in the series only the first loop
contributions\footnote{For combining the total resummation with
NLO DGLAP one more term in the series should be retained}:

\begin{equation}\label{ser}
H_{NS}^{(1)} = \frac{A(\omega C_F)}{2 \pi}
\Big[\frac{1}{\omega}+\frac{1}{2}\Big],~~C_{NS}^{(1)} =1
 + \frac{A(\omega C_F)}{2 \pi}
\Big[\frac{1}{\omega^2}+\frac{1}{2\omega}\Big]~.
\end{equation}
Finally, there is \textbf{Step C:}  Subtract the first-loop
expressions (\ref{ser}) from Eq.~(\ref{sum})) to get the combined,
or "synthetic" as we called them in Ref.~\cite{egtfit},
coefficient function $c_{NS}$ and anomalous dimension $h_{NS}$:
\begin{equation}\label{syn}
c_{NS} = \tilde{c}_{NS} -C_{NS}^{(1)},~~~~~h_{NS} = \tilde{h}_{NS}
-H_{NS}^{(1)}.
\end{equation}
Substituting Eqs.~(\ref{syn}) in Eq.~(\ref{fnsint}) leads to the
expression for $g_1^{NS}$ equally good at large and small $x$.
This description does not require singular factors in the fits for
the initial parton densities. An alternative approach for
combining DLA expression for $g_1$ was suggested in
Ref.~\cite{kwe}. However, the parametrization of $\alpha_s$ in
this approach was simply borrowed from DGLAP, which makes this
approach be unreliable at small $x$.
\section{Conclusion}

We have briefly considered the essence of the IREE method together
with examples of its application to different processes. They
demonstrate that IREE is indeed the  efficient and reliable
instrument for all-orders calculations in QED, QCD and the
Standard Model of EW interactions. As an example in favor of this
point, let us just remind that there exist wrong expressions for
the singlet $g_1$ in DLA obtained with an alternative technique
and the exponentiation of EW double logarithms obtained in
Ref.~\cite{flmm} had previously been denied in several papers
where other methods of all-order summations were used.

\section{Acknowledgement}
B.I.~Ermolaev is grateful to the Organizing Committee of the
Epiphany Conference for financial support of his participation in
the conference.

\end{document}